\documentclass[a4paper,11pt]{article}
\usepackage{pos}

\usepackage{siunitx}
\usepackage{subcaption}
\usepackage{todonotes}
\title{A Combined Fit of the Diffuse Neutrino Spectrum using IceCube Muon Tracks and Cascades}
 \ShortTitle{A Combined Fit of the Diffuse Neutrino Spectrum using IceCube Muon Tracks and Cascades}

\author{The IceCube Collaboration \\{\normalsize \normalfont(a complete list of authors can be found at the end of the proceedings)}}

\emailAdd{erik.ganster@icecube.wisc.edu}
\emailAdd{rnaab@icecube.wisc.edu}
\emailAdd{zelong.zhang@icecube.wisc.edu}

\abstract{The IceCube Neutrino Observatory first observed a diffuse flux of high energy astrophysical neutrinos in 2013. Since then, this observation has been confirmed in multiple detection channels such as high energy starting events, cascades, and through-going muon tracks. Combining these event selections into a high statistics global fit of 10 years of IceCube's neutrino data could strongly improve the understanding of the diffuse astrophysical neutrino flux: challenging or confirming the simple unbroken power-law flux model as well as the astrophysical neutrino flux composition. One key component of such a combined analysis is the consistent modelling of systematic uncertainties of different event selections. This can be achieved using the novel SnowStorm Monte Carlo method which allows constraints to be placed on multiple systematic parameters from a single simulation set. We will report on the status of a new combined analysis of through-going muon tracks and cascades. It is based on a consistent all flavor neutrino signal and background simulation using, for the first time, the SnowStorm method to analyze IceCube's high-energy neutrino data. Estimated sensitivities for the energy spectrum of the diffuse astrophysical neutrino flux will be shown.

\vspace{4mm}
{\bfseries Corresponding authors:}
Erik Ganster$^{1*}$, Richard Naab $^{2}$, Zelong Zhang$^{3}$\\
{$^{1}$ \itshape RWTH Aachen University}\\
{$^{2}$ \itshape DESY Zeuthen}\\
{$^{3}$ \itshape Stony Brook University}\\[4mm]
$^*$ Presenter
} % from abstract

\FullConference{37$^{\rm{th}}$ International Cosmic Ray Conference (ICRC 2021)\\
		July 12th -- 23rd, 2021\\
		Online -- Berlin, Germany}

\begin{document}
\maketitle

%\todo[inline]{This is based on the "official" ICRC2021 PoS template but the authorship/author style is copied from the IceCube internal PoS template. This is based on an old (2019) template which causes some errors/warnings in the log as some functions/control sequences have changed and are not compatible anymore, e.g. \textbackslash speaker}
%\listoftodos
%\clearpage

% section 1
\section{Introduction}
%\begin{itemize}
%    \item IceCube's high energy astrophysical neutrino flux measured in 2013
%    \item Confirmation of observation in multiple detection channels
%    \item GlobalFit from Lars
%    \item Here: MC study of a combination of tracks and cascades using a consistent treatment of systematic uncertainties
%\end{itemize}

Since the discovery of a high energy astrophysical neutrino flux in 2013 \cite{Aartsen:2013jdh}, the IceCube Neutrino Observatory has confirmed the measurement of a diffuse extra-galactic neutrino flux in several detection channels such as: high energy starting events \cite{Abbasi:2020jmh}, through-going muon tracks \cite{Aartsen:2016xlq} and cascades \cite{Aartsen:2020aqd}. The measurements prefer a single power-law (SPL) energy spectrum for the astrophysical neutrino flux, and the measured properties (normalization and spectral index) by these complementary analyses are consistent within uncertainties.

%The measured spectrum of these neutrinos is harder than the atmospheric neutrino spectrum and the arrival directions of these astrophysical neutrinos are consistent with an isotropic distribution, therefore favoring an extra-galactic origin.

% alternative 
%The IceCube experiment has discovered a flux of astrophysical neutrinos in 2013 \cite{}.
%The spectrum of these neutrinos is harder than the atmospheric neutrino spectrum and the arrival directions of the astrophysical neutrinos are consistent with an isotropic distribution, favoring an extra-galactic origin.
%Several other measurements using different detection channels confirmed this observation, and in 2015 a combined analysis of IceCube's neutrino data was performed \cite[Lars]{}.

In 2015, the first combined analysis of IceCube's high energy neutrino data was performed \cite{Aartsen:2015knd}. A new combined analysis targeting the energy spectrum of the diffuse astrophysical neutrino flux and utilizing the higher statistics of the complementary track-like and cascade detection channels in IceCube is currently being prepared.
%Using 10 years of IceCube's high energy neutrino data, this combined analysis allows to challenge the single power-law astrophysical neutrino flux model as well it's flavor composition.
Here, we report on the current status and estimated sensitivity of this combined analysis.
%It relies on consistent modeling of systematic uncertainties for both selections for which a novel Monte Carlo (MC) simulation method is used.
% suggestion for last sentence: 
It relies on consistent modeling of the neutrino flux components and corresponding uncertainties as well as a consistent treatment of detector systematic uncertainties across all measurement channels. We use a novel Monte Carlo (MC) simulation technique and present a new method for including this MC in an analysis.

% Here, we report on the status of a combined analysis utilizing IceCube's track-like and cascade detection channel which is currently beeing prepared. These complementary event selections not only feature a very low overlap of events, making them ideal candidates for a combined fit of IceCube's high energy neutrino data, but also focus on different neutrino flavors/interactions. This allows to challenging the simple unbroken power-law and the astrophysical neutrino flux composition.

%Here, we present a combined analysis utilizing the track and cascade detection channels that is based on a new simulation method for taking detector systematic uncertainties into account.
%We describe this new method and, based on the consistently used MonteCarlo (MC) event simulation, describe the event overlap between the cascade and track channel.
%Finally, we show the expected precision of such a combined analysis for 10 years of data.

% section 2
%\section{The SnowStorm MC}
\section{A novel approach for including detector systematic uncertainties}
\label{sec:snowstorm_mc}
%\begin{itemize}
%    \item SnowStorm simulation method, paper
%    \item Description of the SnowStorm MC
%    \item SnowStorm systematic treatment
%    \item Sketch of re-weighting using a Gaussian
%\end{itemize}

A combination of multiple event selections into a combined fit relies on consistent modeling of not only the signal parameters but also the systematic uncertainties. The analysis presented here is based on the SnowStorm method \cite{Aartsen:2019jcj} for treating systematic uncertainties within each individual event selection (cascades and through-going tracks).

The central element of the SnowStorm simulation method (as presented in \cite{Aartsen:2019jcj}) is the so-called "SnowStorm event ensemble": each event\footnote{For computational reasons a few events are grouped and treated the same.} is simulated with a set of certain nuisance parameter values that are continuously sampled from a distribution defined in advance. The result of this is an ensemble of events, each representing a different combination of nuisance parameters in the pre-defined nuisance parameter phase-space.

% optionall: Different formulation
% Central element of the SnowStorm simulation method (as presented in \cite{Aartsen:2019jcj}) is the so called "SnowStorm event ensemble". In this ensemble, each event\footnote{In reality this applies not for each single event but for a few events which are grouped together for computational reasons.} has been simulated with a different choice of nuisance parameters. The parameter values are chosen according to an individual sampling distribution for each nuisance parameter. The result of this is an ensemble of events, each representing a different combination of nuisance parameters in the pre-defined nuisance parameter phase-space.

\begin{figure}[htbp]
    \centering
    \includegraphics[width=0.7\textwidth]{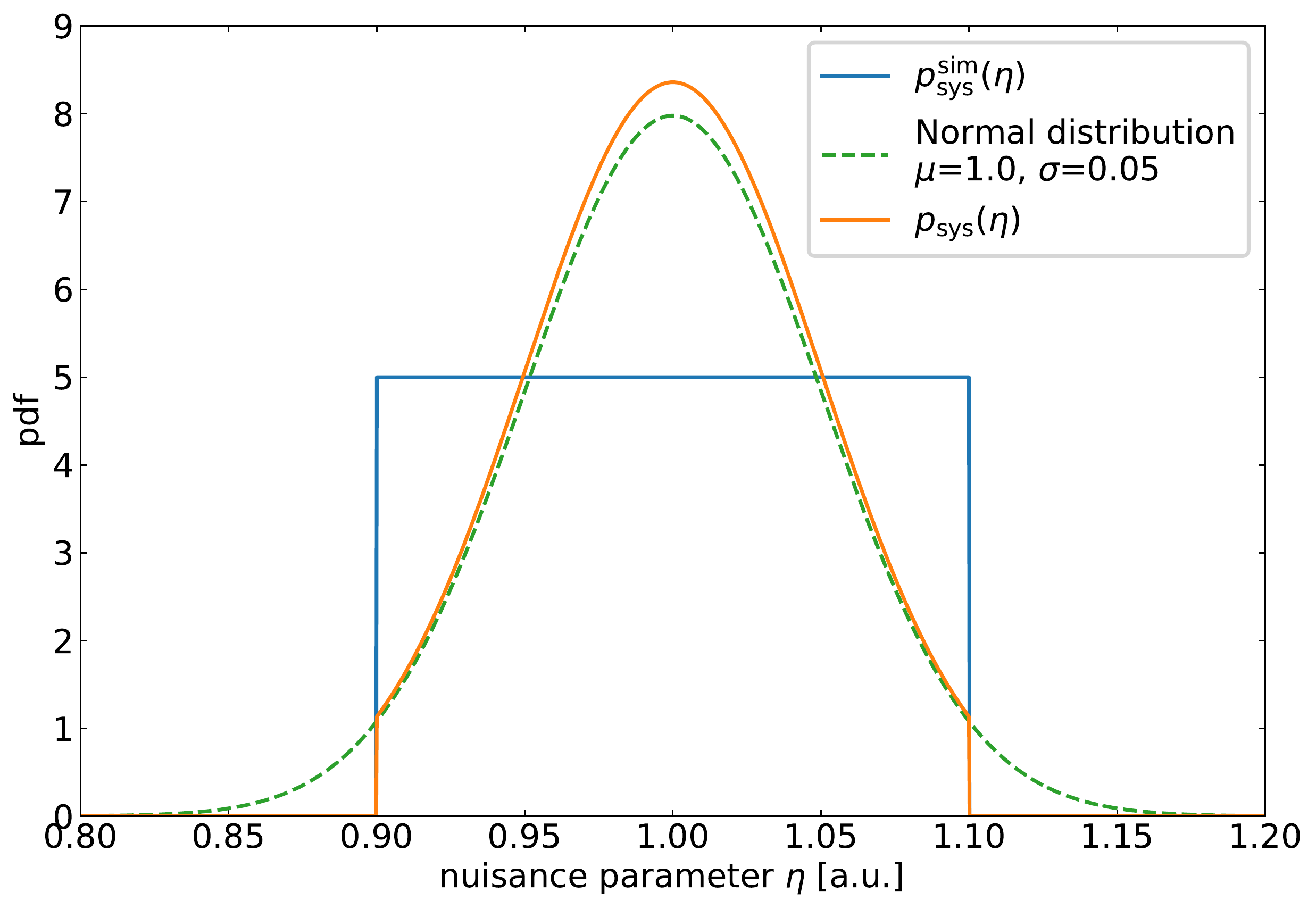}
    \caption{Sketch of re-weighting a SnowStorm event ensemble. The uniformly sampled nuisance parameter $\eta$ (blue), but is re-weighted to a Gaussian/normal distribution for use in the analysis (orange). To account for the simulated phase space, in this case $0.9 < \eta < 1.1$, during simulation, the Gaussian distribution (green) needs to be re-normalized to represent a proper probability density function (orange).}
    \label{fig:reweighting_sketch}
\end{figure}

%In the original method, this event ensemble is used to calculate the gradient of the analysis observables with respect to the nuisance parameters. In combination with a MC set representing the default nuisance parameter value, this gradient is then multiplied with the desired variation in the nuisance parameter to obtain a prediction of the event distribution for different values of the nuisance parameters \cite{Aartsen:2019jcj}. A necessary condition is that second-order variations in the nuisance parameter can be neglected.
%Another possibility is to re-weight the generated event ensemble as shown in Figure \ref{fig:reweighting_sketch}: A nuisance parameter $\eta$, describing a certain systematic uncertainty, has been simulated according to a uniform distribution during generation of the SnowStorm event ensemble $p_{\mathrm{sys}}^{\mathrm{sim}}(\eta)$. To obtain the resulting event distribution in the analysis space for a discrete choice of our nuisance parameter $\eta$, the event ensemble is re-weighted to a Gaussian distribution in $\eta$: $p_{\mathrm{sys}}(\eta) = \mathcal{G}(\eta, \mu, \sigma)$. The mean value $\mu$ is set to the value of $\eta$ that one want to obtain a prediction for, while $\sigma$ can be used to limit the range used for re-weighting. The per event re-weigting factor $w$ is then given by:

The method presented here will re-weight the simulated distribution of events according to the current choice of nuisance parameter values in the fit. This is different from using the event ensemble to obtain the gradient of the analysis observables with respect to the nuisance parameters shown in \cite{Aartsen:2019jcj}. To obtain the event distribution in the analysis space for a discrete choice of some nuisance parameter $\eta$, the SnowStorm event ensemble is re-weighted to a Gaussian distribution in $\eta$: $p_{\mathrm{sys}}(\eta) = \mathcal{G}(\eta, \mu, \sigma)$. The mean value $\mu$ is set to the value of $\eta$ that one wants to obtain a prediction for, while $\sigma$ can be used to limit the range used for re-weighting. The per-event re-weighting factor $w$ is then given by:
\[
%\begin{equation*}
    w(\eta) = {\frac {p_{\mathrm{sys}}(\eta)}{p_{\mathrm{sys}}^{\mathrm{sim}}(\eta)}}
    \,,
%\end{equation*}
\]
where the denominator takes the finite phase space of the sampling distribution $p_{\mathrm{sys}}^{\mathrm{sim}}(\eta)$ used during simulation into account. Assuming the effect of the systematic uncertainties to be sufficiently small so that variations can be treated perturbatively (neglecting $\mathcal{O}(\eta^2)$ terms) as in \cite{Aartsen:2019jcj}, the re-weighted event ensemble will yield the event distribution for a discrete choice of $\eta = \mu$. This re-weighting can be applied for multiple nuisance parameters at the same time by expanding the above calculation to all nuisance parameters $\Vec{\eta} = (\eta_0, \eta_1, \,...\, ,\eta_i)$ for which a re-weighting should be applied:

% optional: Different formulation
%Given the assumption from \cite{Aartsen:2019jcj}, that an integration of the full SnowStorm event ensemble yields the same event distribution as a dedicated baseline simulation set with $\eta = \langle p_{\mathrm{sys}}^{\mathrm{sim}}(\eta) \rangle$ and that $\mathcal{O}(\eta^2)$ terms can be neglected, the re-weighted event ensemble will yield the same event distribution as a simulation set with $\Vec{\eta} = \Vec{\mu}$.

\begin{equation*}
    w(\Vec{\eta}) = \prod_{i} {\frac {p_{\mathrm{sys}}(\eta_i)}{p_{\mathrm{sys}}^{\mathrm{sim}}(\eta_i)}}
    \,.
\end{equation*}

This concept of re-weighting is not limited to a Gaussian re-weighting distribution. Any symmetric $p_{sys}(\eta_i)$ distribution will yield to the prediction of $\eta_i = \langle p_{sys}(\eta_i) \rangle$. Figure \ref{fig:reweighting_sketch} shows the relative contribution to the final analysis event distribution as a function of the nuisance parameter value $\eta$ for the case of $p_{sys} = \mathcal{G}(\eta, \mu, \sigma)$.
For a Gaussian, weights events with a value of $\eta$ close to $\mu$ have a higher weighting factor than events further away.

We have successfully used a SnowStorm neutrino event ensemble with this re-weighting method to take the same systematic uncertainties as used in the individual analyses in \cite{Aartsen:2016xlq} and \cite{Aartsen:2020aqd} (optical efficiency, bulk ice absorption and scattering, and hole ice) into account. Comparisons of this SnowStorm re-weighting treatment with the systematics treatment previously used in the individual analyses showed very similar results.

% optional: Argumentation of Gaussians becoming a delta function in the limit of (infinitesimal) small widths
%The rationale behind this is that in the limit of infinite small $\sigma$ one effectively uses a delta-function $\delta (x-\mu )$, which is equivalent to using a systematic set simulated with a discrete choice of $x=\mu$.

% optional: Different formulations
% However, direct comparison of both SnowStorm methods of dealing with the event ensemble (re-weigthing and gradient extraction) in an analysis are yet to be done.
%Comparisons with the MC and systematic treatment used in previous iterations of the individual standalone analyses show similar results for the sensitivities.

% section 3
\section{Monte Carlo Event Samples and Analysis method}

\subsection{The Event Samples}
%\begin{itemize}
%    \item Short description of track + cascade event selection, references to the papers
%    \item Overlap of both selections
%\end{itemize}

This work aims to combine two of IceCube's neutrino data samples, through-going muon tracks and cascades, into a single analysis. Individual analyses have been performed on both event samples, confirming the observation of a high energy astrophysical neutrino flux \cite{Aartsen:2016xlq, Aartsen:2020aqd}.

The through-going muon track sample used in \cite{Aartsen:2016xlq} and \cite{Stettner:2019tok} focuses on up-going track-like events with a reconstructed zenith angle $\theta_{\mathrm{reco}} > \SI{85}{\degree}$. This cut uses the Earth as a shield against the background of atmospheric muons reaching the IceCube in-ice neutrino detector. This background is further reduced by a boosted decision tree (BDT) trained to separate atmospheric muons from muons originating from charged current muon-neutrino interactions. The result is a high purity (\SI{99.7}{\percent}) sample of muon neutrinos of either atmospheric or astrophysical origin \cite{Aartsen:2016xlq}.

The cascade data sample used in \cite{Aartsen:2020aqd} is a full-sky sample and consists of three sub-samples, cascade signal sample, muon control sample and hybrid sample \cite{thesis:Hans}. It selects on low level cascade events in low and high energy regimes and is classified into three sub-samples by the low energy event selection. The low energy event selection mainly uses a BDT method \cite{thesis:Hans} but the high energy ($E_\mathrm{reco} > \SI{60}{\tera\eV}$) event selection uses straight cuts \cite{thesis:Yiqian}. The cascade signal sample is dominated by conventional atmospheric neutrinos and astrophysical neutrinos. The muon background only contributes $\sim$ 8\% to it \cite{thesis:Hans}. The events are binned into 3 zenith bins, corresponding to northern, horizontal and southern sky and 22 energy bins from $10^{2.6}$ to $10^{7}$ GeV \cite{thesis:Hans}. Since the astrophysical neutrino spectrum fitted in single power law is harder than atmospheric neutrino spectrum, this sample is dominated by atmospheric neutrinos at low energy but astrophysical neutrinos at high energy. The muon control sample is not binned. It is dominated by atmospheric muons ($\sim$ 65\%) and it is used to constrain the normalization of atmospheric muon part in cascade analysis \cite{thesis:Hans}. The hybrid sample is binned into 11 energy bins. It is dominated by conventional atmospheric muon neutrinos interacting via the charged current channel, so it is used to constrain the atmospheric neutrino part in original cascade analysis \cite{thesis:Hans}. 

%
% subsection
%
\subsection{Analysis Method}
\label{sec:analysis_method}
%\begin{itemize}
%    \item LLH combination
%    \item Fit/Nuisance parameters
%    \item Table of fit parameters?
%\end{itemize}

%We use a binned likelihood, to analyze two-dimensional histograms of both event selection, constructed in reconstructed energy and zenith ($E_{\mathrm{reco}}$ and $\mathrm{cos(\theta_{reco})}$). The binning of the histograms of the tracks and cascade event selection is chosen as in the individual analyses. Since we created disjoint event samples by removing the event overlap, the combined likelihood is the product over all now independent bins $ \mathcal{L_\mathrm{combined}} = \prod_{\mathrm{bins \, i}} \mathcal{L}_i $ and is a function of the signal and nuisance parameters described in the following.

Both individual analyses use a two dimensional Likelihood fit of reconstructed energy and zenith ($E_{\mathrm{reco}}$ and $\mathrm{cos(\theta_{reco})}$) for analyzing the data. In the case of fully disjunct samples, a combined Likelihood can be obtained by building the product of the, in this case independent, per-bin Likelihoods:

\begin{equation*}
    \mathcal{L_\mathrm{combined}} = \prod_{\mathrm{bin}\,i}^{N_{\mathrm{bins}}} \mathcal{L}_i \left( n_i, \mu_i(\vec{\rho}, \vec{\eta}) \right)
    %\, \times \prod_j \mathrm{exp}( \frac{(\eta_j-\hat{\eta_j})^2}{2 \sigma_{\eta_j}^2})
    \, \times \prod_j \pi(\eta_j)
    \,,
\end{equation*}

where the number of events $n_i$ in analysis histogram bin $i$ is compared to the expected number of events $\mu_i$ using a Poisson likelihood $\mathcal{L}_i$. The expectation $\mu_i$ is a function of the signal ($\vec{\rho}$) and nuisance ($\vec{\eta}$) parameters. Priors $\pi(\eta_j)$ on the nuisance parameters $\eta_j$ are chosen as described in Table \ref{tab:fit_parameters}.

\begin{table}[htbp]
    \centering
    \begin{tabular}{lccc}
        Name & & Allowed Range & Prior \\
        \hline
        \hline
        spectral index & $\gamma_{\mathrm{astro}}$ & $[1.0, \infty)$ & - \\
        flux normalization & $\Phi_0^{\mathrm{astro}}$ & $[0.0, \infty)$ & -\\
        \hline
        Conventional Flux Normalization & & $[0.0, \infty)$ & -\\
        Prompt Flux Normalization & & $[0.0, \infty)$ & - \\
        Muon Flux Normalization (cascades only) & & $[0.0, \infty)$ & -\\
        Cosmic-Ray model interpolation & $\lambda_{\mathrm{CRModel}}$ & $[-1.0, +2.0]$ & $\mathcal{G}(0.0, 1.0)$\\ 
        Cosmic-Ray spectral index shift & $\Delta_{\gamma_{\mathrm{CR}}}$ & $[-1.0, +1.0]$ & -\\
        \hline
        Optical Efficiency & & $[0.9, 1.1]$ & -\\
        Bulk Ice Absorption & & $[0.9, 1.1]$ & -\\
        Bulk Ice Scattering & & $[0.9, 1.1]$ & -\\
        Hole-Ice $p_0$ & & $[-1.0, +1.0]$ & -\\
        \hline
    \end{tabular}
    \caption{All used fit parameters, their allowed ranges, and Gaussian priors $\mathcal{G}(\mu, \sigma)$ (if used). The horizontal lines separate the signal parameters from the flux and detector nuisance parameters (from top to bottom).}
    \label{tab:fit_parameters}
\end{table}

This is only valid if both individual analysis histograms have no common\footnote{and thus "overlapping"} events. Despite focusing on two different event topologies in the detector, there is some overlap between IceCube's through-going track and cascade event selection. For the Cascade signal and muon sample, there is only a marginal overlap (\SI{0.1}{\percent}) with the tracks sample. This can easily be eliminated by assigning a tag to those overlapping events and using them only once when constructing the analysis histograms.

However, as the hybrid sub-selection of the cascade sample is optimized for starting events with a track-like signature in the detector, about \SI{30}{\percent} of these events also pass the through-going track selection. In the individual analysis of IceCube's cascade data, the starting track sub-sample is used to constrain the normalization of the atmospheric neutrino flux \cite{thesis:Hans}. In the combined fit presented here, the large statistics of the tracks sample allows an even tighter constraint of the atmospheric neutrino flux normalization. Therefore, we do not include the cascade hybrid selection in this combined analysis. We fit the tracks sample as well as the cascade signal and cascade muon control sample.

\subsection{Fit parameters}

We model the astrophysical neutrino component with a SPL in the form:

\begin{align}
    \Phi_{\nu + \bar{\nu}}^{\mathrm{astro}} = c_{\mathrm{units}} \times \Phi_0^{\mathrm{astro}} \times \left( \frac{E_\nu}{\SI{100}{\tera\eV}} \right)^{-\gamma_{\mathrm{astro}}}
    \,,
\end{align}

with the flux normalization $\Phi_0^{\mathrm{astro}}$ and the spectral index $\gamma_{\mathrm{astro}}$ as two free parameters in units of $c_{\mathrm{units}} = 10^{-18} \, \si{\per\GeV \per\cm\squared \per\s \per\steradian}$. We further assume an equal flux of all neutrino flavors as well as neutrinos and antineutrinos. The shape of the spectrum of both conventional and prompt atmospheric neutrino fluxes is obtained using the Matrix-CascadeEquation solver package MCEq \cite{Fedynitch:2015}, and the normalizations of both are left floating.
%We allow for slight spectral distortions ($\Delta_{\gamma_{\mathrm{CR}}}$) of the primary cosmic ray model, as well as an interpolation between the two primary cosmic ray models H4a and GST4 (as in the previous iteration of the tracks analysis \cite{Stettner:2019tok}). 
We allow for slight spectral distortions of the primary cosmic ray model, as well as an interpolation between the two primary cosmic ray models H4a and GST4 (as in the previous iteration of the tracks analysis \cite{Stettner:2019tok}). An independent normalization of the flux of atmospheric muons contributing to the cascade event sample is also left floating in the fit.

Finally, the likelihood also depends on some detector systematic parameters, which are treated using the re-weighting method of the used SnowStorm event ensemble as explained in \ref{sec:snowstorm_mc}. In agreement with the latest standalone analyses, we include variations in the optical efficiency, the absorption and scattering coefficients of the bulk ice as well as variations in the parameter controlling for effects of the refrozen hole-ice \cite{Aartsen:2020aqd, Stettner:2019tok}. Using the re-weighting method presented here, we obtain very similar results compared to the systematic treatment previously used in the individual analyses.

%Applying the method of re-weighting to the individual sub-analyses allows for a comparison to the MC and systematic treatment used in previous iterations. Figure \ref{fig:2d_astro_SPL} shows that the re-weighting of the SnowStorm event ensemble yields very similar sensitivity results. It is therefore used in the following sensitivity estimate of the combined analysis of IceCube's cascade and track data described here.
%\todo{We do not show the comparison to the previous tracks result anymore. Explain in words that with SnowStorm we achieve very similar results as with discrete systematics sets?}

%For comparing the method of re-weighting the SnwoStorm event ensemble with the MC and systematic treatment used in both previous analyses, we have performed fits using only the cascade/through-going track selection.

% section 4
\section{Results/Sensitivities}
\label{sec:results}
%\begin{itemize}
%    \item Estimated sensitivity
%    \item comparison to standalone analyses
%    \item "piece-wise SPL unfolding" ?
%\end{itemize}

\begin{figure}[htbp]
    \centering
    \includegraphics[width=0.75\textwidth]{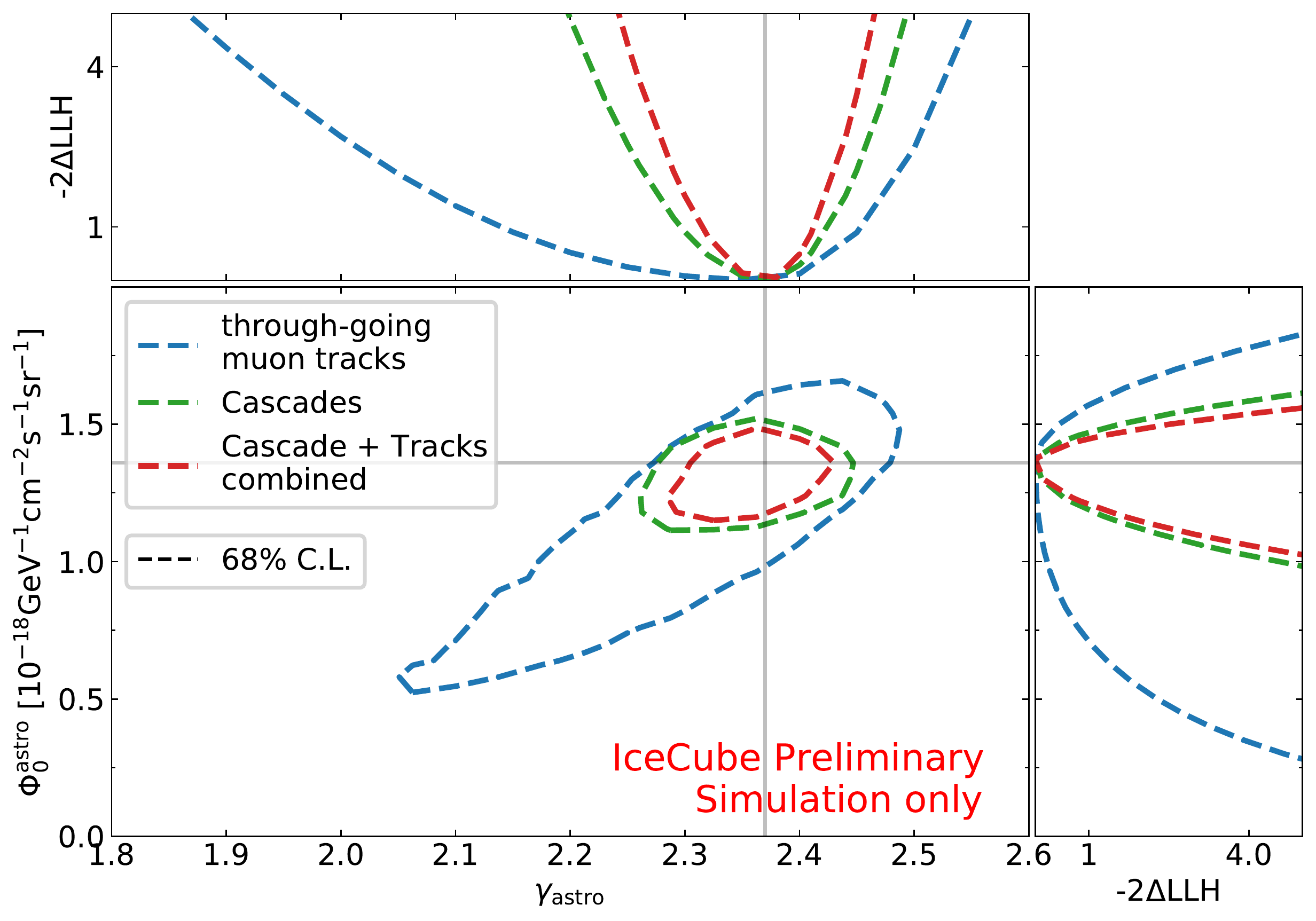}
    \caption{68\% CL contours of the signal parameters $\Phi_{\mathrm{0}}^{\mathrm{astro}}$ and $\gamma_{\mathrm{astro}}$. The red line shows the fit of combining track and cascade (Asimov) data, individual fit results are shown in blue (tracks) and green (cascades). For all contours, a single power-law flux was used for generating the Asimov signal and in the hypothesis used for fitting. Grey lines show the parameter values of the generated Asimov signal. The top and right plots show 1D profile Likelihood scans of both signal parameters.}
    \label{fig:2d_astro_SPL}
\end{figure}

Table \ref{tab:fit_results_SPL} lists the estimated sensitivities of the combined fit. For the Asimov signal, the best fit of \cite{Stettner:2019tok} was assumed. We further assumed a lifetime of 10 years for all fits. The estimated 68\% CL contours for the astrophysical signal parameters are shown in Figure \ref{fig:2d_astro_SPL}. The sensitivities for the individual through-going track and cascade analysis, assuming the same Asimov signal and lifetime, are shown for comparison. Compared to these, the sensitivity of the combined analysis of both event selections is strongly increased.

% old text
%Figure \ref{fig:2d_astro_SPL} shows the estimated sensitivity of the combined analysis. For the Asimov signal, the best fit of \cite{Stettner:2019tok} was assumed. We estimate that the signal parameters can be measured as $\Phi_{\mathrm{0}}^{\mathrm{astro}} = 1.36^{+0.09}_{-0.15}$ and $\gamma_{\mathrm{astro}} = 2.37^{+0.04}_{-0.06}$ assuming lifetime of 10 years. The estimated \SI{90}{\percent} CL contours for the astrophysical signal parameters resulting from the individual tracks and cascades analysis assuming the same 10 year lifetime and using the same detector systematic treatment are shown for comparison.
%For the same Asimov signal and lifetime, the individual analysis yield sensitivities of $\Phi_{\mathrm{0}}^{\mathrm{astro}} = 1.36^{+0.21}_{-0.65}$ and $\gamma_{\mathrm{astro}} = 2.37^{+0.08}_{-0.23}$ for the through-going muon tracks and $\Phi_{\mathrm{0}}^{\mathrm{astro}} = 1.36^{+0.11}_{-0.17}$ and $\gamma_{\mathrm{astro}} = 2.37^{+0.06}_{-0.07}$ for the cascades.

\begin{table}[htbp]
    \centering
    \begin{tabular}{lccc}
        & through-going tracks & cascades & combined fit \\
        \hline
         flux normalization $\Phi_0^{\mathrm{astro}}$ & $1.36^{+0.21}_{-0.65}$ & $ 1.36^{+0.11}_{-0.17}$ & $1.36^{+0.09}_{-0.15}$ \\
        \rule{0pt}{3ex} spectral index $\gamma_{\mathrm{astro}}$ & $2.37^{+0.08}_{-0.23}$ & $2.37^{+0.05}_{-0.07}$ & $2.37^{+0.04}_{-0.05}$ \\
        \hline
    \end{tabular}
    \caption{Estimated 68\% sensitivities for an Asimov signal fit of an injected single power-law astrophysical neutrino flux.}
    \label{tab:fit_results_SPL}
\end{table}

The smaller size of the contours for cascades compared to the tracks results from the much better energy resolution of shower-type events (mainly selected by cascade selection) compared to track events: Whereas in the case of cascades, a "calorimetric" measurement of the energy is possible, the measurement of the energy of a muon only provides a lower limit to the parent neutrino's energy when the muon was produced outside the detector. On the other hand, the large effective area of the track selection leads to a much higher rate of observed events which constrains the nuisance parameters of the atmospheric fluxes and detector systematic uncertainties much more as in an analysis targeting cascades only.

As explained in Section \ref{sec:analysis_method}, a decision has to be made about the overlapping events: They can be used either in the tracks sample or in the cascade sample(s). What we find is that the measurement of the astrophysical signal parameters is independent of the decision made. Given the small number of overlapping events in total as well as the fact that starting tracks do not constitute the actual signal for the cascade sample, this is not surprising.

\begin{figure}[htbp]
    \centering
    \includegraphics[width=0.8\textwidth]{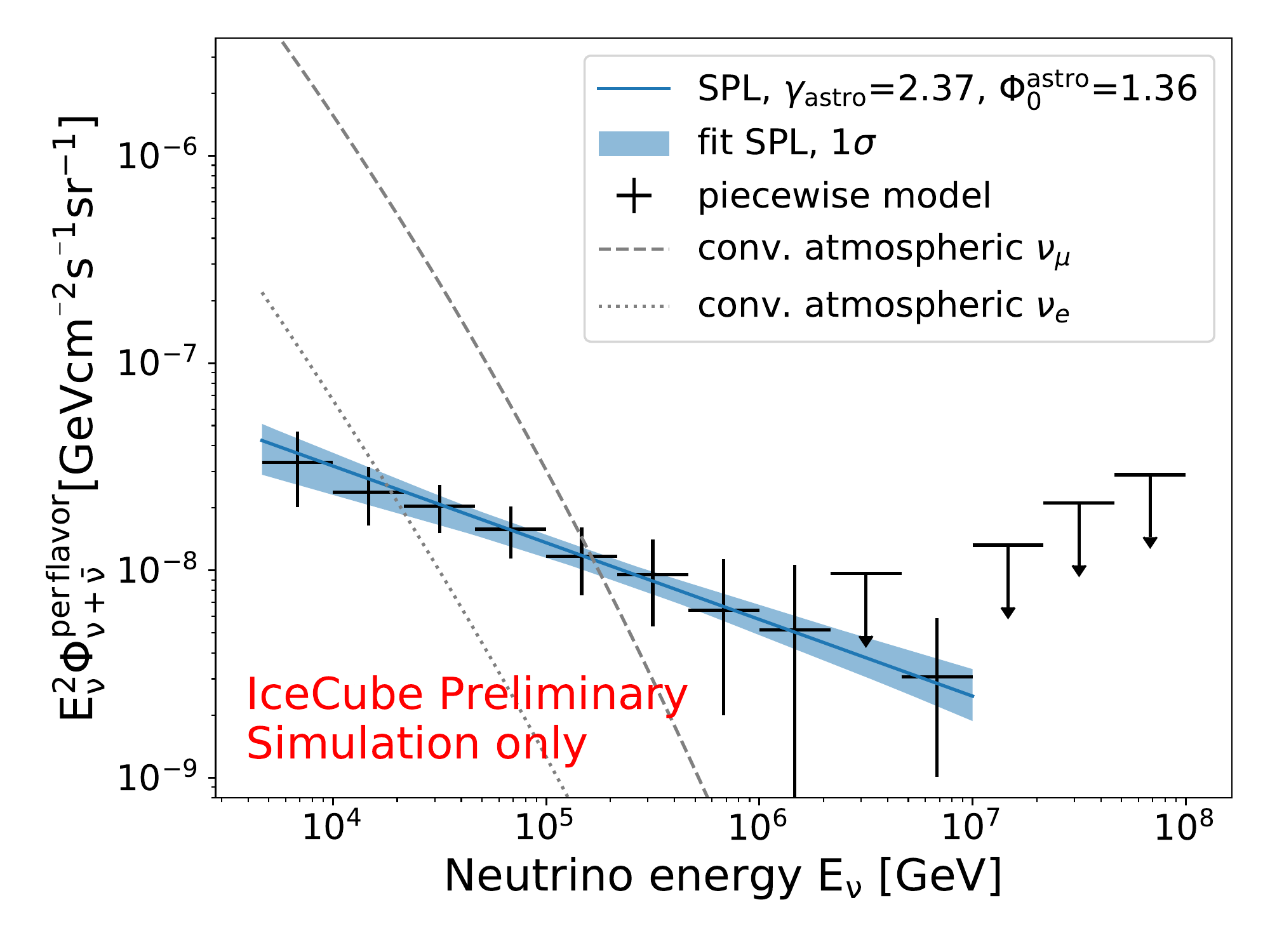}
    \caption{Inferred neutrino spectrum from the combined analysis. The blue line represents the SPL model assumed for the study presented here. The blue region shows the estimated $1\sigma$ CL when fitting a SPL model. Alternatively, a piecewise model describing the flux with independent contributions from different bins in energy with a per-bin spectral index of -2, is fit and the results are shown in black crosses. In the highest energy segment without an upper limit, the sensitivity to astrophysical neutrinos is enhanced due to the Glashow resonance \cite{IceCube:2021rpz}.}
    \label{fig:piecewise_unfolding}
\end{figure}

Figure \ref{fig:piecewise_unfolding} shows the spectrum inferred from the combined fit, along with the result of fitting a piecewise model to the assumed signal. The piecewise model assumes independent normalizations in each segment of neutrino energy, with a fixed spectral index of $\gamma_{\mathrm{astro}}=-2$ in every energy bin. The flux level of atmospheric neutrinos is shown for comparison. In our combined analysis, the flux is determined most precisely around 100 TeV. At higher energies, where atmospheric backgrounds are low, we expect a low number of events overall, which limits the analysis. At lower energies, the astrophysical flux is subdominant, limiting the precision to which it can be measured.
In the case of down-going events, atmospheric neutrinos can be accompanied by muons produced in the same air shower. These muons are vetoed in the cascade event selection so that atmospheric neutrinos have a certain "self-veto" probability to be discarded by the selection criteria. The suppression factor can be $\gtrapprox10$ for very down-going neutrinos at 10 TeV \cite{thesis:Hans}, so that the sensitivity to the astrophysical flux in the corresponding zenith analysis bins extends to lower energies.
%The atmospheric self-veto effect for downward-going events, relevant for the cascade sample, alleviates this problem to some extent.

% section 5
\section{Conclusions and Outlook}

We have presented the current status of a combined diffuse fit of IceCube's high energy cascade and through-going muon track neutrino data. This study makes use of a SnowStorm MC event ensemble and a new method for analyzing this ensemble as described in Section \ref{sec:snowstorm_mc}. This provides a consistent treatment of the detector systematic nuisance parameters for both event selections. Along with using MCEq for obtaining the atmospheric neutrino contributions, this allows for a combination of the cascade and through-going muon track event selections into a single Likelihood analysis.

The sensitivities presented in Section \ref{sec:results} assume a livetime of 10 years and will allow to challenge the so far single-power law modeled astrophysical neutrino flux. The benefits of combining different analyses were demonstrated by comparing our result to the individual analyses using exactly the same modeling. The approach presented here can be easily extended to included further detection channels as well as additional uncertainties on the atmospheric fluxes or the modeling of the ice. This will be investigated further in the future.

%in this sense, the sensitivities demonstrate the benefits of combining different analyses.
%In the future, we want to study the analysis performance considering additional uncertainties on the atmospheric neutrino fluxes ("Barr" paper?/Jöran's ICRC paper?) and more systematic uncertainties related to the imperfect knowledge of the glacial ice (anisotropy). Finally, including more detection channels like for example starting tracks will hopefully improve this combined analysis.

%% Use IceCube's internal ICRC bibtex style
\bibliographystyle{ICRC}
\bibliography{references}

%\begin{thebibliography}{99}
%\bibitem{...}
%....
%
%\end{thebibliography}

% Full authors list (ONLY FOR COLLABORATIONS)
\clearpage
% Full authors list (ONLY FOR COLLABORATIONS)
%\clearpage
\section*{Full Author List: IceCube Collaboration}

% \noindent \textbf{Note comment afterwards:} Collaborations have the possibility to provide an authors list in xml format which will be used while generating the DOI entries making the full authors list searchable in databases like Inspire HEP. For instructions please go to icrc2021.desy.de/proceedings or contact us under icrc2021proc@desy.de.\\

% \scriptsize
% \noindent
% first.author$^1$, 
% second.author$^2$, 
% third.author$^3$ % .... more names
% and 
% last.author$^{n}$ \\

% \noindent
% $^1$first.affiliation.
% $^2$second.affiliation. % .... more affiliation
% $^{m}$last.affiliation.

\scriptsize
\noindent
R. Abbasi$^{17}$,
M. Ackermann$^{59}$,
J. Adams$^{18}$,
J. A. Aguilar$^{12}$,
M. Ahlers$^{22}$,
M. Ahrens$^{50}$,
C. Alispach$^{28}$,
A. A. Alves Jr.$^{31}$,
N. M. Amin$^{42}$,
R. An$^{14}$
% ...
\\

\noindent
% ...
$^{12}$ Universit{\'e} Libre de Bruxelles, Science Faculty CP230, B-1050 Brussels, Belgium \\
% ...
$^{14}$ Department of Physics and Laboratory for Particle Physics and Cosmology, Harvard University, Cambridge, MA 02138, USA \\
% ...
$^{17}$ Department of Physics, Loyola University Chicago, Chicago, IL 60660, USA \\
$^{18}$ Dept. of Physics and Astronomy, University of Canterbury, Private Bag 4800, Christchurch, New Zealand \\
% ...
$^{22}$ Niels Bohr Institute, University of Copenhagen, DK-2100 Copenhagen, Denmark \\
% ...
$^{28}$ D{\'e}partement de physique nucl{\'e}aire et corpusculaire, Universit{\'e} de Gen{\`e}ve, CH-1211 Gen{\`e}ve, Switzerland \\
% ...
$^{31}$ Karlsruhe Institute of Technology, Institute for Astroparticle Physics, D-76021 Karlsruhe, Germany  \\
% ...
$^{42}$ Bartol Research Institute and Dept. of Physics and Astronomy, University of Delaware, Newark, DE 19716, USA \\
% ...
$^{50}$ Oskar Klein Centre and Dept. of Physics, Stockholm University, SE-10691 Stockholm, Sweden \\
% ...
$^{59}$ DESY, D-15738 Zeuthen, Germany \\
% ...

\subsection*{Acknowledgements}

\noindent
USA {\textendash} U.S. National Science Foundation-Office of Polar Programs,
U.S. National Science Foundation-Physics Division,
U.S. National Science Foundation-EPSCoR,
Wisconsin Alumni Research Foundation,
Center for High Throughput Computing (CHTC) at the University of Wisconsin{\textendash}Madison,
Open Science Grid (OSG),
Extreme Science and Engineering Discovery Environment (XSEDE),
Frontera computing project at the Texas Advanced Computing Center,
U.S. Department of Energy-National Energy Research Scientific Computing Center,
Particle astrophysics research computing center at the University of Maryland,
Institute for Cyber-Enabled Research at Michigan State University,
and Astroparticle physics computational facility at Marquette University;
Belgium {\textendash} Funds for Scientific Research (FRS-FNRS and FWO),
FWO Odysseus and Big Science programmes,
and Belgian Federal Science Policy Office (Belspo);
Germany {\textendash} Bundesministerium f{\"u}r Bildung und Forschung (BMBF),
Deutsche Forschungsgemeinschaft (DFG),
Helmholtz Alliance for Astroparticle Physics (HAP),
Initiative and Networking Fund of the Helmholtz Association,
Deutsches Elektronen Synchrotron (DESY),
and High Performance Computing cluster of the RWTH Aachen;
Sweden {\textendash} Swedish Research Council,
Swedish Polar Research Secretariat,
Swedish National Infrastructure for Computing (SNIC),
and Knut and Alice Wallenberg Foundation;
Australia {\textendash} Australian Research Council;
Canada {\textendash} Natural Sciences and Engineering Research Council of Canada,
Calcul Qu{\'e}bec, Compute Ontario, Canada Foundation for Innovation, WestGrid, and Compute Canada;
Denmark {\textendash} Villum Fonden and Carlsberg Foundation;
New Zealand {\textendash} Marsden Fund;
Japan {\textendash} Japan Society for Promotion of Science (JSPS)
and Institute for Global Prominent Research (IGPR) of Chiba University;
Korea {\textendash} National Research Foundation of Korea (NRF);
Switzerland {\textendash} Swiss National Science Foundation (SNSF);
United Kingdom {\textendash} Department of Physics, University of Oxford.
%\section*{Full Authors List: \Coll\ Collaboration}
%
%\noindent \textbf{Note comment afterwards:} Collaborations have the possibility to provide an authors list in xml format which will be used while generating the DOI entries making the full authors list searchable in databases like Inspire HEP. For instructions please go to icrc2021.desy.de/proceedings or contact us under icrc2021proc@desy.de.\\
%
%\scriptsize
%\noindent
%first.author$^1$, 
%second.author$^2$, 
%third.author$^3$ % .... more names
%and 
%last.author$^{n}$ \\
%
%\noindent
%$^1$first.affiliation.
%$^2$second.affiliation. % .... more affiliation
%$^{m}$last.affiliation.

\end{document}